\documentclass[aps,prl,twocolumn,superscriptaddress,showpacs]{revtex4-1}
\usepackage{amsmath,amssymb,graphics,epsfig,epstopdf,color,verbatim,ulem,braket,tabularx}
\usepackage[colorlinks,linkcolor=blue,citecolor=blue,urlcolor=blue]{hyperref}

\graphicspath{{Figures/}}

\begin{document}

\title{Reaching the continuum limit in finite-temperature {\it ab initio} field-theory computations in many-fermion systems}

\author{Yuan-Yao He}
\address{Center for Computational Quantum Physics, Flatiron Institute, New York, New York 10010, USA}
\address{Department of Physics, College of William and Mary, Williamsburg, Virginia 23187, USA}
\author{Hao Shi}
\address{Center for Computational Quantum Physics, Flatiron Institute, New York, New York 10010, USA}
\author{Shiwei Zhang}
\address{Center for Computational Quantum Physics, Flatiron Institute, New York, New York 10010, USA}
\address{Department of Physics, College of William and Mary, Williamsburg, Virginia 23187, USA}

\begin{abstract}
Finite-temperature, grand-canonical computations based on field theory are widely
applied in areas including condensed matter physics, ultracold atomic gas systems, and lattice gauge theory. However, these calculations have computational costs scaling as $N_s^3$ with the size of the lattice or basis set, $N_s$. We report a new approach based on systematically controllable low-rank factorization which reduces the scaling of such computations to $N_s N_e^2$, where $N_e$ is the average number of fermions in the system. In any realistic calculations aiming to describe the continuum limit, $N_s/N_e$ is large and needs to be extrapolated effectively to infinity for convergence. The method thus fundamentally changes the prospect for finite-temperature many-body computations in correlated fermion systems. Its application, in combination with frameworks to control the sign or phase problem as needed, will provide a powerful tool in {\it ab initio} quantum chemistry and correlated electron materials. We demonstrate the method by computing exact properties of the two-dimensional Fermi gas with zero-range attractive interaction, as a function of temperature in both the normal and superfluid states.
\end{abstract}


\date{\today}
\maketitle

Computations are playing an increasingly important role in addressing the fundamental challenges of understanding strong correlations in interacting quantum systems.
Understandably, a major part of the development and application efforts in both physics and chemistry have focused on ground-state properties. However, experimental conditions are always at finite temperatures, where often rich and new properties can be revealed~\cite{Damascelli2003,Fischer2007,Mueller2017,Brown2019}. One example is the rapid development in the area of experiments with ultracold atoms, where temperature plays a crucial role and very precise measurements of properties are often possible with exquisite control over interaction strengths, environments, etc~\cite{Bloch2008,Leticia2018}. Accurate computations of thermodynamic properties allow direct comparisons with experiments, but are challenging because of the presence of strong coupling and thermal fluctuations. A second example is in strongly correlated materials, including high-temperature superconductors~\cite{Bednorz1986,Saleem2013},
where some of the outstanding and most interesting physics questions concern finite-temperature properties~\cite{Patrick2006}.

A common finite-temperature formalism is based on field theory in which the thermodynamic properties are computed as path-integrals in field space. Approximations can be used to perform the path integrals, including the simplest, namely mean-field calculations. A more powerful approach is to evaluate the many-dimensional integration by Monte Carlo methods~\cite{Blankenbecler1981}. This has become a key technique for many-body finite-temperature computations, widely applied in several fields of physics~\cite{Scalapino1981,Hirsch1983,Hirsch1985,ALHASSID2001,Chandrasekharan2006,Chandrasekharan2010,Chandrasekharan2017,ALHASSID2010,ALHASSID2016,Alhassid2018,Jensen2019,Anderson2015}. For example, many of the sign-problem-free computations in lattice models in condensed matter, and in Fermi gas and optical lattices of ultracold atoms, have been performed this way. For general Hamiltonians, for instance the doped Hubbard model or realistic electronic Hamiltonians in solids or molecules, a constraint can be applied to control the sign or phase problem. This framework, even with simple trial wave functions to impose an approximate constraint, has been shown to be very accurate and has been applied widely in ground-state calculations~\cite{Shiwei1995,Shiwei1997,Zhang1997,Carlson1999,Shiwei2003,Suewattana2007,Chang2010,Mingpu2016b,Vitali2019}. Finite-temperature generalization of the approach has also been developed~\cite{Shiwei1999,Shiwei2000,Liu2018,Yuanyao2019}.

These finite temperature calculations all have computational complexity of $\mathcal{O}(N_s^3)$~\cite{LOH2005}, as they are formulated in the grand-canonical ensemble to analytically evaluate the fermion trace along each path in auxiliary-field space, leading to determinants with dimension $N_s$. In the majority of applications,
for example dilute Fermi gas and all {\it ab initio\/} real material simulations, it is necessary to reach the continuum (large lattice or complete basis set) limit in order to obtain realistic results. One must take $N_s\rightarrow\infty$, while keeping the average number of fermions $N_e$ fixed (at the targeted number of electrons in the molecule or cluster, or in the supercell). In contrast, all ground-state calculations, which are formulated in canonical ensemble and only need to retain occupied orbitals along the path in field space, scale as $\mathcal{O}(N_sN_e^2)$~\cite{Shihao2015}. The ratio $N_s/N_e$ is often $\mathcal{O}(10$-$100)$ or larger for realistic calculations. The discrepancy of $(N_s/N_e)^2$ in computational cost can render calculations inaccessible at low, or even modest, temperatures, when the corresponding ground-state calculations can be performed straightforwardly. This poses a fundamental obstacle for finite-temperature studies in interacting fermion systems.

We address this problem in the present paper, introducing a new method which allows finite-temperature computations to be performed with complexity $\mathcal{O}(N_sN_e^2)$, with no loss of accuracy. We demonstrate the approach within the framework of sign-problem-free determinantal quantum Monte Carlo (DQMC) calculations, in the two-dimensional (2D) dilute Fermi gas with contact interaction, which has attracted intense experimental interest. The algorithm yields speedups of several orders of magnitude over the standard approach, and allows computations of exact thermodynamic properties in the continuum. This adds a new dimension in our computational repertoire; for example, it allows direct {\it ab initio} computations of the superfluid state in strongly interacting Fermi gases and investigations of the corresponding phase transitions. 

For concreteness in describing the algorithm and to specify the computational details for our results on the 2D dilute Fermi gas, we use the attractive Hubbard model on a square lattice:
\begin{equation}
\label{eq:HubbardModel}
\begin{split}
\hat{H} = \sum_{\mathbf{k}\sigma}\varepsilon_{\mathbf{k}}c_{\mathbf{k}\sigma}^+c_{\mathbf{k}\sigma}
-\mu\sum_{\mathbf{i},\sigma} \hat{n}_{\mathbf{i}\sigma}
+U\sum_{\mathbf{i}}\hat{n}_{\mathbf{i}\uparrow}\hat{n}_{\mathbf{i}\downarrow}\,,
\end{split}
\end{equation}
where  $\varepsilon_{\mathbf{k}}=2t(2-\cos k_x-\cos k_y)$, $\sigma$ ($=\uparrow$ or $\downarrow$) denotes spin, and $\hat{n}_{\mathbf{i}\sigma}=c_{\mathbf{i}\sigma}^+c_{\mathbf{i}\sigma}$ is the density operator. The periodic supercell is represented by a lattice of size $N_s=L^2$, with corresponding momentum $k_x$ (and $k_y$) defined in units of $2\pi/L$. The energy scale of the system is set by $t$. The chemical potential $\mu/t$ is tuned to target the desired number of particles in the supercell, $N_e=N_{\uparrow}+N_{\downarrow}$, defining a lattice density $n=N_e/N_s$. The interaction strength $U/t$ is uniquely determined by $\log(k_Fa)$~\cite{Shihao2015}, i.e., the ratio of the two-particle scattering length to the average interparticle spacing (given by the  Fermi wave vector $k_F$). We will measure energies in units of the Fermi energy $E_F=2\pi n t$ and temperatures in units of $T_F\equiv E_F/k_B$. 

\begin{figure}[t]
\centering
\includegraphics[width=0.80\columnwidth]{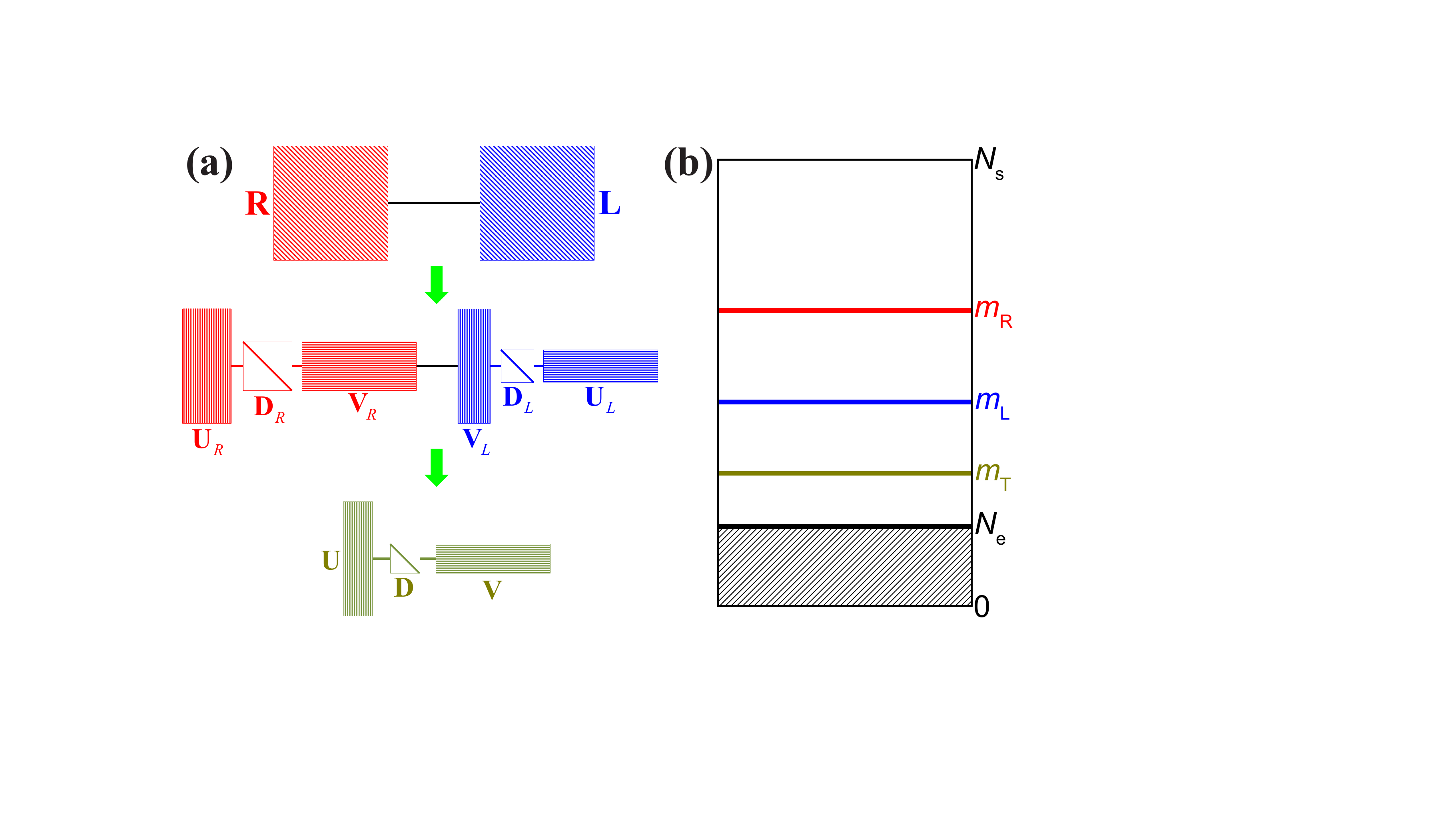}
\caption{\label{fig:Demonstration} Schematic illustration of the decomposition and truncations along a field path. The path is separated at $\tau$ into two parts, $\mathbf{L}$ and $\mathbf{R}$, which are $N_s\times N_s$ matrices of propagator products [top row of (a)]. They are factorized and truncated with controlling dimensions $m_{\text{L}}$ and $m_\text{R}$, respectively, which change dynamically as $\tau$ evolves in the sampling [(b) and second row of (a)]. A further factorization can be performed after $\mathbf{L}$ and $\mathbf{R}$ are combined [bottom row in (a)], leading to a further truncation with dimension $m_\text{T}$.}
\end{figure}

The partition function $Z=\text{Tr}(e^{-\beta\hat{H}})$, after imaginary-time discretization $\beta=\Delta\tau M$, Trotter decomposition and Hubbard-Stratonovich (HS) transformation, can be written as~\cite{Yuanyao2019}
\begin{equation}
\label{eq:Partition}
Z \simeq \int P(\mathbf{X})\prod_{\sigma={\uparrow,\downarrow}}\det(\mathbf{1}_{N_s}+\mathbf{B}_M^{\sigma}\mathbf{B}_{M-1}^{\sigma}\cdots\mathbf{B}_2^{\sigma}\mathbf{B}_1^{\sigma})\,d{\mathbf{X}}\,,
\end{equation}
where $\mathbf{X}=\{\mathbf{x}_1,\mathbf{x}_2,\cdots,\mathbf{x}_M\}$ is a path in auxiliary-field space, and $P(\mathbf{X})$ is a probability density function. The auxiliary fields at each time slice, $\mathbf{x}_{\ell}=\{x_{i,\ell}\}$, can be either continuous (e.g., $P(\mathbf{X})$ is a Gaussian) or discrete fields (as we adopt below for our Fermi gas calculations), in which case the integral becomes a sum.
The "$\simeq$" indicates the Trotter error, which can be extrapolated away
with calculations using smaller values of $\Delta\tau$. The one-body propagator $\mathbf{B}_{\ell}^{\sigma}$ depends on the auxiliary-field $\mathbf{x}_{\ell}$, and is an $N_s\times N_s$ matrix. The paths $\mathbf{X}$ are sampled according to the integrand in Eq.~(\ref{eq:Partition}), for example by Metropolis Monte Carlo in standard DQMC, or by a branching random walk of fixed length $M$ in constrained-path auxiliary-field quantum Monte Carlo (AFQMC)~\cite{Yuanyao2019}.

To describe our algorithm, we introduce the notation $\mathbf{R}=\mathbf{B}^{\sigma}_{\ell}\cdots\mathbf{B}^{\sigma}_2\mathbf{B}^{\sigma}_1$ and $\mathbf{L}=\mathbf{B}^{\sigma}_{M}\mathbf{B}^{\sigma}_{M-1}\cdots\mathbf{B}^{\sigma}_{\ell+1}$, where $\ell$ is an arbitrary time slice along the path. The key components of DQMC and AFQMC calculations include: propagation; numerical stabilization; evaluations of ratios or the derivative of the integrand with respect to $\mathbf{x}_{\ell}$ during updates; and measurements. These procedures all involve operations of the matrix products $\mathbf{L}$ and  $\mathbf{R}$. For example, update and measurements at $\tau=\ell\Delta\tau$ require the equal-time single-particle Green's function matrix $\mathbf{G}^{\sigma}(\tau,\tau)\equiv \{\langle c_{i\sigma}c^+_{j\sigma}\rangle_{\tau}\}=(\mathbf{1}_{N_s}+\mathbf{R}\mathbf{L})^{-1}$. Because of the manipulation of these matrices and the computation of the determinant, the overall computational complexity scales as $\mathcal{O}(N_s^3)$~\cite{LOH2005} (with a proportionality factor containing the imaginary-time length $M$). This basic structure is depicted in the top row in Fig.~\ref{fig:Demonstration}(a).

\begin{figure*}
\centering
\includegraphics[width=0.90\textwidth]{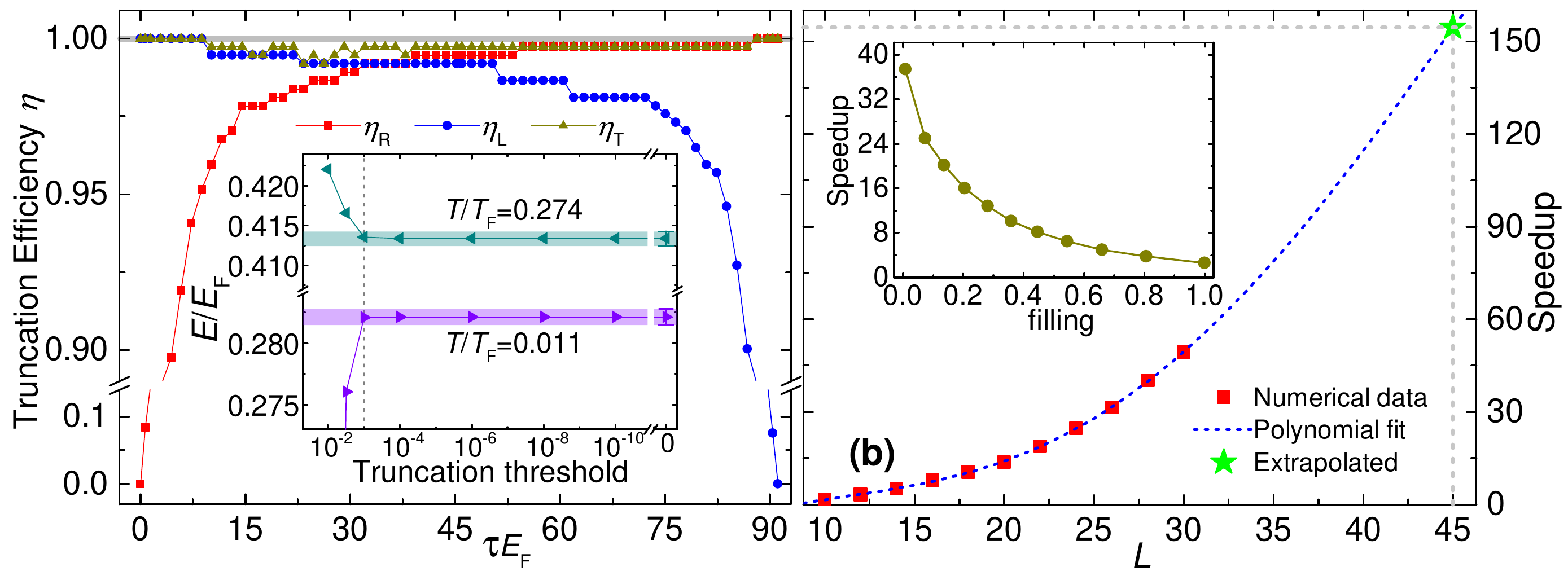}
\caption{\label{fig:SpeedErr} Efficiency and accuracy of the factorization and truncation, and computational scaling and speedup. In (a), truncation efficiency $\eta_{\text{L},\text{R},\text{T}}$ (see text) is shown for a system with $L=20$, $N_e=58$, $U/t=-3$ and $T/T_F=0.011$. Note zoom in vertical scale for $\eta>0.9$. The inset shows the residual error of the computed total energy per particle vs.~truncation threshold for two temperatures, $T/T_F=0.274$ ($\beta t=4$) and $T/T_F=0.011$ ($\beta t=100$). Full result is shown on the right as zero truncation threshold, with typical final statistical errors indicated by the shade. In (b), the main graph shows speedup vs.~lattice linear dimension for $\log(k_Fa)=4.346574$, $N_e=58$ and $T/T_F=1/32$, while the inset presents speedup vs.~lattice density for $L=32$, $U/t=-3$ and $\beta t=32$.}
\end{figure*}

The matrices can be written in factorized forms, for example using the column-pivoted QR algorithm to perform $\mathbf{U}\mathbf{D}\mathbf{V}$ decompositions (see e.g., Ref.~\onlinecite{Yuanyao2019}), to obtain: $\mathbf{R}=\mathbf{U}_R\mathbf{D}_R\mathbf{V}_R$ and $\mathbf{L}=\mathbf{V}_L\mathbf{D}_L\mathbf{U}_L$, where $\mathbf{D}_R,\mathbf{D}_L$ are diagonal matrices with positive elements in descending order and $\mathbf{U}_R,\mathbf{U}_L,\mathbf{V}_R,\mathbf{V}_L$ are $N_s\times N_s$ matrices, which are used in standard methods as part of the procedure to keep numerical stability ~\cite{White1989,LOH2005,Yuanyao2019}. The diagonal elements of $\mathbf{D}_L$ and $\mathbf{D}_R$ typically span an enormous scale (e.g., $10^{+100}\sim10^{-100}$), controlled by the independent-particle spectrum, interaction strength, and inverse temperature. We now introduce a threshold $\epsilon$ to perform truncations on them in computing $\mathbf{R}$ and $\mathbf{L}$, since contributions from the elements of $\mathbf{D}_R$ and  $\mathbf{D}_L$ smaller than $\epsilon$ will be bounded and can be made smaller than numerical noise if the truncation threshold $\epsilon$ is sufficiently small. Thus, $\epsilon$ is a parameter that we can tune to control the numerical precision of the calculation. Suppose the number of elements in $\mathbf{D}_R$ larger than $\epsilon$ is $m_{\text{R}}$, then $\mathbf{U}_R$,$\mathbf{D}_R$,$\mathbf{V}_R$ are effectively $N_s\times m_{\text{R}}$, $m_{\text{R}}\times m_{\text{R}}$, $m_{\text{R}}\times N_s$ matrices. Similarly, we have a truncation dimension $m_{\text{L}}$ for $\mathbf{D}_L$, and low-rank approximation for $\mathbf{L}$. The truncations are illustrated in Fig.~\ref{fig:Demonstration}. We can then decompose $\mathbf{R}\mathbf{L}=\mathbf{U}\mathbf{D}\mathbf{V}$ and an additional truncation on $\mathbf{D}$ can be performed, with truncation dimension $m_{\text{T}}$, as shown in the bottom row of Fig.~\ref{fig:Demonstration}(a). We should note that, methodologically, the above idea fits into a theme of low-rank factorizations which have found broad applications in very different contexts in physics and chemistry (e.g.,density matrix renormalization group~\cite{White1992,*White1993} and tensor hyper contraction of quantum chemical Hamiltonians~\cite{Hohenstein2012,*Parrish2012}).

With the new formalism, updating all the auxiliary fields on a single imaginary-time slice costs $\mathcal{O}(N_sm_{\text{T}}^2)$ to calculate the force bias~\cite{Shihao2015} and $\mathcal{O}(m_{\text{T}}^3)$ to calculate the ratio of the determinant in Eq.~(\ref{eq:Partition}), applying $\det(\mathbf{1}_{N_s}+\mathbf{U}\mathbf{D}\mathbf{V})=\det(\mathbf{1}_{m_{\text{T}}}+\mathbf{D}\mathbf{V}\mathbf{U})$. Propagation along the path, i.e., moving $\ell$ to the left or right, becomes a propagation on $\mathbf{U}_R$ and $\mathbf{U}_L$, which can be achieved with $\mathcal{O}(N_sm_{\text{R},\text{L}}\log N_s)$ scaling using FFT and locality of the interaction in Eq.~(\ref{eq:HubbardModel}). For a single numerical stabilization, the complexity is also lowered to $\mathcal{O}(N_sm_{\text{R},\text{L}}^2)$, since we are dealing with $\mathbf{U}_R$ and $\mathbf{U}_L$ matrices. For measurements, we compute the Green's function as $\mathbf{G}^{\sigma}(\tau,\tau)=(\mathbf{1}_{N_s}+\mathbf{U}\mathbf{D}\mathbf{V})^{-1}=\mathbf{1}_{N_s}-\mathbf{U}\mathbf{D}(\mathbf{1}_{m_{\text{T}}}+\mathbf{V}\mathbf{D}\mathbf{U})^{-1}\mathbf{V}$.

In the calculations, $m_{\text{R}}$, $m_{\text{L}}$ and $m_{\text{T}}$ evolve dynamically, as we carry out the truncations following each numerical stabilization procedure. At the start and end of a sweep, $m_{\text{R}}$ or $m_{\text{L}}$ is close to $N_s$, but away from the ends they rapidly decay, as illustrated in Fig.~\ref{fig:SpeedErr}(a). The ``truncation efficiency'', defined as $\eta_{\text{L},\text{R},\text{T}}=(N_s-m_{\text{L},\text{R},\text{T}})/(N_s-N_{\sigma})$, remains over 90\% except for small portions at the ends. The computational complexity is dominated by the main part of the path in the middle, which scales as $\mathcal{O}(N_sm_{\text{T}}^2)$, and the overall $m_{\text{T}}$ is small across the entire path.  In standard DQMC, lowering the temperature makes the calculation progressively more challenging; as $T\rightarrow 0$, the computational efficiency reaches its worst in comparison with the ground-state method. In our new method, $m_{\text{T}}$ approaches $N_e$ as $T$ is lowered, and the algorithm restores the same computational complexity as the ground-state projection approach. At higher temperatures $m_{\text{T}}$ increases. In the example in Fig.~\ref{fig:SpeedErr}(a) $m_{\text{T}}$ is $\sim1.3\times$ and $\sim2.5\times$ the ground-state value at $T/T_F=0.055$ and $0.274$, respectively, but even in the latter a speedup of $\sim 30$ is achieved. As mentioned earlier, the closer to the continuum limit, the larger the speedup, since $m_{\text{T}}$ only depends on physical parameters such as $N_e$ and $T$, and will change little as $N_s$ is increased. 

As shown in the inset in Fig.~\ref{fig:SpeedErr}(a), the truncation error is invisible even with an aggressive truncation threshold of $\epsilon=10^{-3}$. (The comparison of different $\epsilon$ vs.~$\epsilon=0$, i.e.~the standard algorithm, was done on an identical set of random paths. In practical calculations the different choices of $\epsilon$ would cause the calculations to, in sufficiently long runs, cease being correlated in their random number streams, but they will give statistically compatible results.) We tested that these results are not sensitive to the temperature. We have also carried out simulations for different interaction strengths ($U/t=-1$ and $-6$) and obtained similar results. Moreover, these results hold generally for different HS transformations, e.g., total density or spin-$s_z$ channels for Eq.~(\ref{eq:HubbardModel})~\cite{Hirsch1983}.

The speedup of our new algorithm over current state-of-the-art is illustrated in Fig.~\ref{fig:SpeedErr}(b). The main figure displays the speedup vs.~the linear dimension of the lattice for 2D Fermi gas system. The speedups fit well a fourth-order polynomial for $L$ up to $30$, consistent with the earlier conclusion of $(N_s/m_{\text{T}})^2$ from computational complexity. The speedup is around $50$ at $L=30$, and is extrapolated to $\sim 150$ for $L=45$, the data point for which is obtained for the new algorithm with five days' computing (on 40 Skylake cores) and which would have required over two years' computing with the standard algorithm using comparable resources. The inset illustrates the speedup vs.~filling by tuning the chemical potential in the context of the Hubbard model. Large speedups are seen at low density, as expected. Even at half-filling, a speedup of $2.65$ is achieved.

\begin{figure}[b]
\includegraphics[width=0.80\columnwidth]{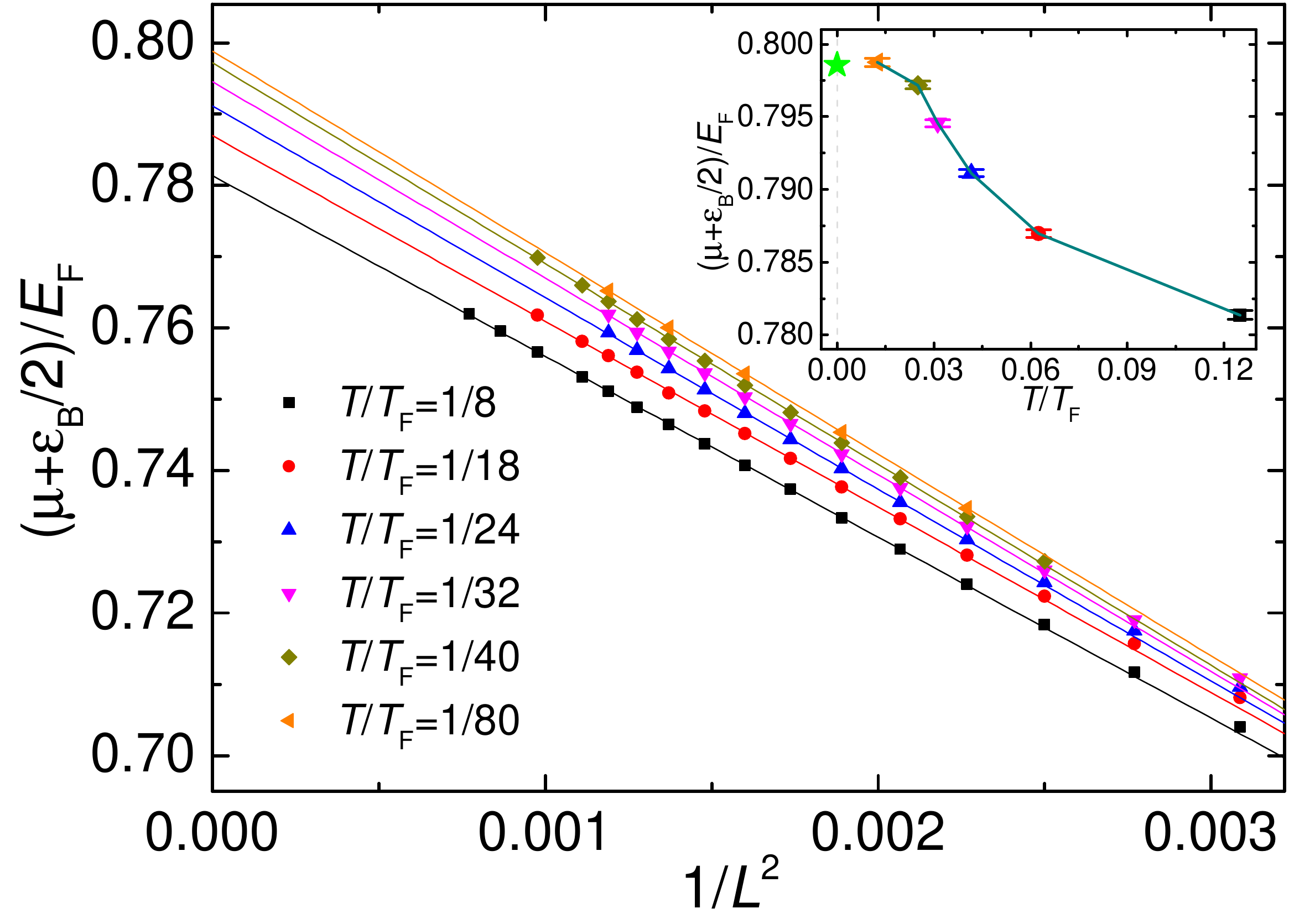}
\caption{\label{fig:EquOfState} Equation of state calculated at $\log(k_Fa)=4.346574$ with $N_e=58$ as a function of temperature. The main graph plots $(\mu+\varepsilon_B/2)/E_F$ versus square of inverse lattice size, and linear fits in the asymptotic regime. Numerical uncertainties are smaller than the symbol size. The inset presents the corresponding results at the continuum limit, which converges to the ground state result (green pentagram, from Ref.~\onlinecite{Shihao2015}) with decreasing temperature. }
\end{figure}

\begin{figure}[t]
\includegraphics[width=0.80\columnwidth]{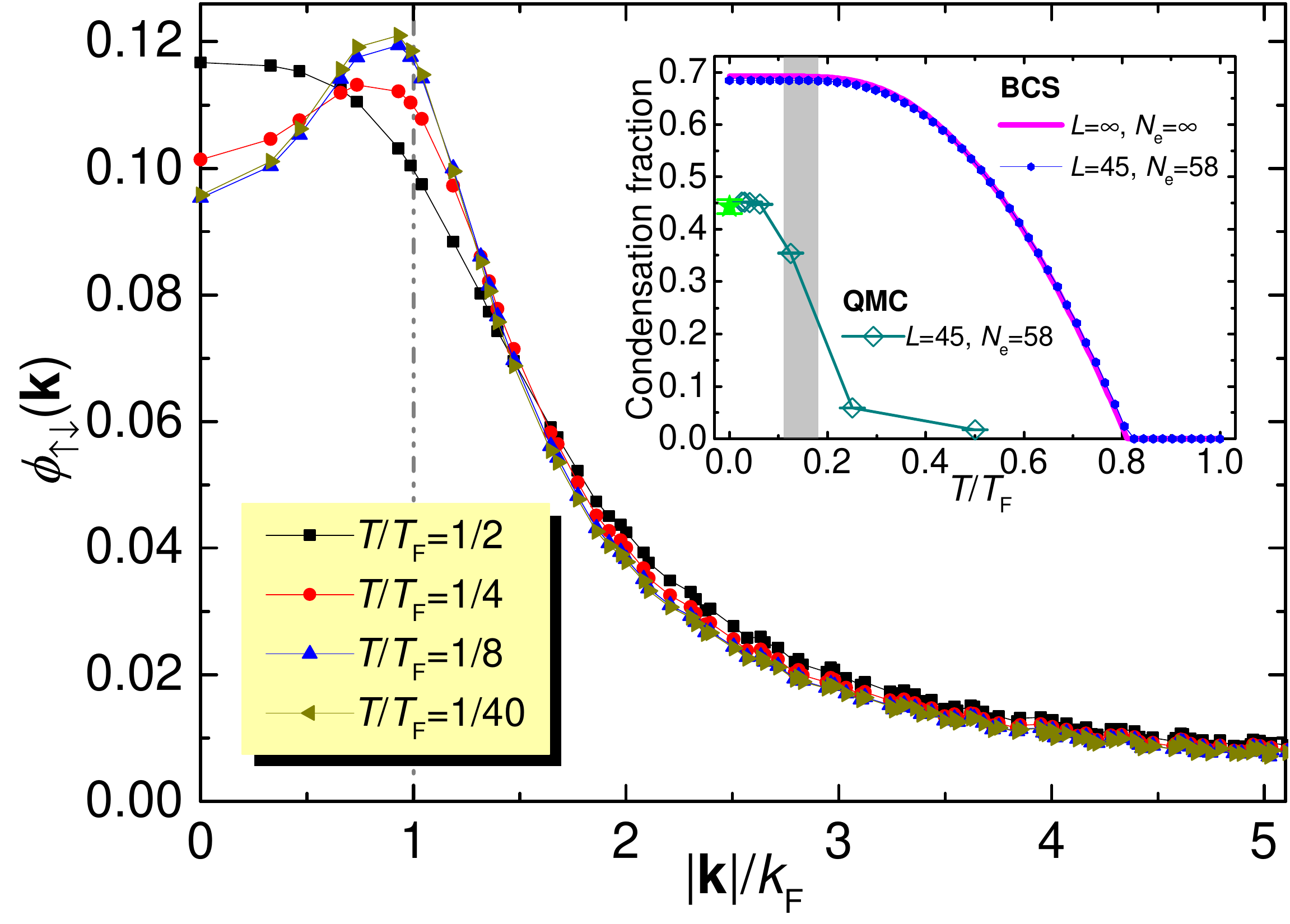}
\caption{\label{fig:SingletPair} Spin-singlet paring wavefunction in reciprocal space versus $|\mathbf{k}|/k_F$, and condensate fraction, for $\log(k_Fa)=0.50$ with $L=45,N_e=58$, as a function of temperature. In the main graph, the error bars in $\phi_{\uparrow\downarrow}(\mathbf{k})$ are smaller than the symbol size. The inset shows the condensate fraction from both QMC and mean-field calculations, and the exact ground state  result (green pentagram, from Ref.~\onlinecite{Shihao2015}). Experimental $T_c/T_F$ for the BKT transition (from Ref.~\onlinecite{Ries2015}) is also shown in the inset with the gray shade indicating uncertainty.}
\end{figure}

We next apply the method to the spin-balanced uniform 2D Fermi gas with a zero-range attractive interaction. Major experimental efforts are being devoted to this and related systems~\cite{Martiyanov2010,Ries2015,Murthy2015,Boettcher2016,Fenech2016,Hueck2017,Murthy452}, which offer opportunities for highly controllable and clean experiments in a strongly interacting 2D quantum system. To our knowledge, finite-$T$ {\it ab initio\/} quantum Monte Carlo computations to date have been mostly limited to the normal state~\cite{Anderson2015,Mulkerin2017}. The algorithmic advances presented in this paper allow exact computations to reach much larger lattices and lower $T$'s, which adds a new dimension to our computational and theoretical capabilities, and will help calibrate and guide experiments. Here, as an illustration of the method, we present examples on the equation of state (EOS) and  pairing properties. 

In Fig.~\ref{fig:EquOfState}, we present the results of EOS  at the interaction strength $\log(k_Fa)=4.346574$, which is on the BCS side. As shown in the main plot, 
the chemical potential shifted by the bound-state energy, $(\mu+\varepsilon_B/2)/E_F$ [note $\varepsilon_B=4\hbar^2/ma^2e^{2\gamma}$ with Euler's constant $\gamma=0.57721$], has perfect linear scaling with $1/L^2$ at large $L$ for all the temperatures. After extrapolation to the continuum limit, $(\mu+\varepsilon_B/2)/E_F$ converges to the the ground-state result when $T\rightarrow 0$, as shown in the inset. Our results indicate that the small discrepancy between experiment~\cite{Boettcher2016} and ground-state results~\cite{Shihao2015} is consistent with finite-$T$ effects, which can now be better understood as experimental resolution improves.

We have also determined the pairing wave function and condensate fraction as a function of temperature. We compute the zero-momentum spin-singlet pairing matrix: $\mathbf{M}_{\mathbf{k}\mathbf{k}^{\prime}}=\langle\Delta_{\mathbf{k}}^+\Delta_{\mathbf{k}^{\prime}}\rangle-\delta_{\mathbf{k}\mathbf{k}^{\prime}}\langle c_{\mathbf{k}\uparrow}^+c_{\mathbf{k}\uparrow}\rangle\langle c_{-\mathbf{k}\downarrow}^+c_{-\mathbf{k \downarrow}}\rangle$ with $\Delta_{\mathbf{k}}^+=c_{\mathbf{k}\uparrow}^+c_{-\mathbf{k}\downarrow}^+$. Its leading eigenvalue divided by $N_e/2$ is identified as the condensate fraction. The corresponding eigenvector gives the pairing wavefunction $\phi_{\uparrow\downarrow}(\mathbf{k})$. Results are shown in Fig.~\ref{fig:SingletPair} for $\log(k_Fa)=0.50$, in the strongly interacting crossover regime. The mean-field results in the inset illustrate the strong correlation effect. The small difference between the finite-size and thermodynamic limit results also help provide a gauge of the finite-size effects in the many-body results. At high temperatures, the pairing wave function is more extended in momentum space, and the condensate fraction is tiny. As the temperature lowers, the pairing wave function peaks more at the Fermi surface, and the condensate fraction increases rapidly. The Berezinskii-Kosterlitz-Thouless (BKT) transition temperature measured experimentally~\cite{Ries2015} is also shown in the inset of Fig.~\ref{fig:SingletPair}. It is clear that the new method now makes possible a quantitative comparison with experiments. A more systematic study of the physics of the 2D Fermi gas and the BKT transition will be published separately.

We have described the algorithm with some specificity in the context of DQMC for a lattice model. As mentioned, the advances are general and can be applied to many forms of finite-temperature field-theory computations. For example, the inclusion of spin-flip terms in the effective single-particle Hamiltonian in Eq.~(\ref{eq:Partition}), such as spin-orbit coupling, can be accommodated straightforwardly~\cite{Shihao2016PRL,Rosenberg2017}. In the presence of a sign problem, e.g., repulsive or spin-imbalanced Hubbard-like models, the constrained path AFQMC framework~\cite{Shiwei1999,Shiwei2000,Liu2018,Yuanyao2019} can be used. In that case $\mathbf{L}$ becomes the trial propagator matrix $\mathbf{L}_T$, and only ${\mathbf R}$ depends on the path, which is active only up to the current $\ell$ and is ``grown'' stochastically to full length. The low-rank factorizations can be applied in exactly the same way. For electronic Hamiltonians with long-range interactions as in molecules and solids, a phase problem arises and the phaseless approximation~\cite{Shiwei2003,Liu2018} is needed. The low-rank factorizations and the rest of the algorithm are identical to the sign problem case. (For a plane-wave basis~\cite{Suewattana2007}, one can alternate between Fourier and real space~\cite{Suewattana2007} similar to the Hubbard model and the computational scaling is similar to what we have discussed. For generic basis sets such as in quantum chemistry, however, the $\mathbf{B}$'s have a more general structure and the propagation scales as $\mathcal{O}(N_s^2 m_{\rm L,R})$, which means the reduction in computational complexity is one power of $N_s/N_e$.) Beyond fermion systems, the factorization and truncation also apply to bosons and Fermi-Bose mixtures~\cite{Rubenstein2012}. More broadly, the idea should also be applicable to calculations employing similar formalisms in nuclear shell models~\cite{ALHASSID2010,ALHASSID2016} and lattice QCD~\cite{Chandrasekharan2003,Chandrasekharan2006,Chandrasekharan2007,Bazavov2010,Petreczky2012}.

In summary, we have presented a method to perform finite-temperature field-theoretic calculations of many-fermion systems with computational complexity $\mathcal{O}(N_s N_e^2)$, i.e,, linear in lattice (or single-particle basis) size. Such calculations are applied widely, and their fundamental complexity of $\mathcal{O}(N_s^3)$ previously has been a major obstacle for approaching the continuum limit ($N_s/N_e\rightarrow \infty$). 
We demonstrated our new method in 2D strongly interacting Fermi gas, where the continuum limit can be reached via numerically exact calculations with large lattice sizes even at low temperatures. The method introduced here can be applied to a variety of strongly interacting fermion systems in multiple fields, including ultracold atomic gases, condensed matter and materials, and quantum chemistry.

We thank J.~Carlson, P.~Dumitrescu and S.~Chandrasekharan for helpful discussions.
This work was supported by NSF (Grant No. DMR-1409510). The Flatiron Institute is a division of the Simons Foundation.

\bibliography{FTQMCScalingMain}

\end{document}